\documentclass[12pt]{article}
\usepackage{epsfig}
\usepackage{axodraw}
\usepackage{amsfonts}

\newcommand{\Li}{\mbox{Li}}

\begin{document}

\begin{center}
{\Large \bf Pion form factor and QCD sum rules: \\ case of pseudoscalar current.}
\\ \vspace*{5mm} V.V.Braguta$^{a}$ and A.I.Onishchenko$^{b,c}$
\end{center}

\begin{center}
a) Institute for High Energy Physics, Protvino, Russia \\
\vspace*{0.5cm} 
b) Department of Physics and Astronomy \\ Wayne State University,
Detroit, MI 48201, USA \\
\vspace*{0.5cm} 
c) Institute for Theoretical and Experimental
Physics, \\ Moscow, Russia 
\end{center}

\vspace*{0.5cm}

\abstract{
We present an analysis of QCD sum rules for pion form factor in next-to-leading
order of perturbation theory for the case of pseudoscalar pion currents. 
The essential instanton contribution is reanalysed with account for present more
accurate values of parameters entering Single Instanton Approximation (SIA). 
The theoretical curve obtained for $Q^2$ dependence of pion form factor is in a good agreement with
existing experimental data. To calculate NLO corrections for double spectral densities
we developed an effective computational technic. The details of the method together
with the results for pion form factor in a more theoretically clean case of axial interpolating
currents will be presented elsewhere.
}\\

\section{Introduction}

One of the first applications, following the discovery of the fact, that perturbation theory can be applied to study
exclusive high momentum transfer processes \cite{Lepage:1980,Efremov:1979,Chernyak:fk}, was the estimates 
of asymptotic $Q^2\to\infty$ behavior of electromagnetic form factors of hadrons. It turned out, that for moderate 
momentum transfers the agreement between asymptotic QCD formulas and experimental data for pion and proton form 
factors is very poor. The latter is not surprising, if one recalls, that the validity of pQCD predictions depends 
in a large extent on the possibility to neglect power corrections ${\cal O}(Q^{-N})$, where $Q$ is the momentum 
transferred, and leave only leading contribution. For pion the leading $1/Q^2$-contribution in the region 
$Q^2\to\infty$ is given by hard rescattering of valence quarks (one gluon exchange subprocess). But, a straightforward 
use of the asymptotic formalism in the region $Q^2\leq 10~\mbox{GeV}^2$ leads to the conclusion, that the mean virtuality 
of the exchanged gluon is smaller then $300$ MeV. Such contributions should be accounted for as pion wave function 
corrections, which receive additional power suppression and at large $Q^2$ scale only as $1/Q^4$. 
Also, in this region finite size or nonperturbative effects become important and as a consequence, a 
complete evaluation of all contributions within pQCD approach utilizing light-cone wave functions at moderate values of 
momentum transfer is impossible.

First systematic study of finite size corrections to asymptotic pQCD formulas was performed in \cite{Botts:kf,Li:1992nu}.
It was concluded, that an account for transverse momenta of quarks as well as Sudakov corrections makes
pQCD predictions self-consistent at momentum transfers as low as few GeV. Moreover, it was argued that pQCD
prediction dominated form factors even at moderate values of momentum transfer. However, a more careful 
analysis \cite{Jakob:1993iw}, specifically taking into account an "intrinsic" pion transverse wave function,
lead to the conclusion, that even self-consistency of pQCD predictions was achieved, it is still insufficient
to describe experimental data.     
 
On the other hand, an alternative approach \cite{Ioffe:ia,Ioffe:qb,Nesterenko:1982gc,Nesterenko:1984tk} based on 
direct evaluation of form-factors from QCD sum rules easily accounts for experimentally measured values
of pion electromagnetic form factor $F_{\pi}(Q^2)$. In this approach a leading order result at moderate
values of $Q^2$ is given by soft wave function overlap mechanism and scales as $1/Q^4$ at large momentum transfers.
Since then, the importance of inclusion both hard (pQCD) and soft contributions to describe experimental data 
was further emphasized and discussed. Later, in this paper, we will focus on QCD sum rule approach, which 
significantly reduces model dependence of the results obtained. So, let us discuss different contributions to 
pion electromagnetic form factor in the language of QCD sum rules.   

The method of QCD sum rules \cite{QCDSR} is designed to estimate low-energy characteristics of hadrons,
such as masses, decay constants and form factors. Within this framework we can easily go into the region of
intermediate momentum transfers and take due care about nonperturbative effects. 

It is already become a standard\footnote{See, for example, \cite{hardsoft}}, when considering the region of moderate 
momentum transfers for pion form factor, to account both for hard scattering and soft wave function overlap mechanism. 
Within QCD sum rule approach the soft contribution is dual to the lowest-order diagram, while the hard contribution
is given by diagrams having higher order in coupling constant $\alpha_s$ and as a consequence suppressed
relative to soft contribution with additional factor $\alpha_s/\pi\sim 0.1$. This extra suppression 
is in a complete agreement with asymptotic behavior of the pion electromagnetic form factor, calculated
in pQCD \cite{Lepage:1980,Efremov:1979,Chernyak:fk}:
\begin{eqnarray}
F_{\pi}^{\mathbf{hard}}(Q^2) = \frac{8\pi\alpha_s (Q^2)}{9}\int_0^1 dx \int_0^1 dy
\frac{\phi_{\pi}(x)\phi_{\pi}(y)}{xyQ^2} = \frac{8\pi\alpha_s f_{\pi}^2}{Q^2}
\end{eqnarray}
At asymptotically high $Q^2$ the ${\cal O}(\alpha_s/\pi)$ suppression of hard contribution is
more than compensated by its slower decrease with $Q^2$. However, such  compensation does not
occur in the region of moderate momentum transfer, where, as we have seen, the soft contribution
may be as important as hard contribution. 

The pion electromagnetic form factor was studied using a bunch of different frameworks, like
QCD sum rules \cite{Ioffe:ia,Ioffe:qb,Nesterenko:1982gc,Nesterenko:1984tk}, light-cone sum rules
\cite{Braun:ij,Braun:1999uj,Bijnens:2002mg}, and different pQCD based approaches employing convolution with
phenomenological pion distribution amplitudes. It should be mentioned that within all these approaches
correlation functions with axial interpolating currents were studied. Another possibility
is to consider a correlation functions with pseudoscalar currents, used as pion interpolating
currents\cite{Forkel:1994pf,Faccioli:2002jd}. This simple change of pion interpolating current has however 
crucial consequences for the effects responsible for deviation of pion electromagnetic form factor from pQCD predictions. 
It is known, that in scalar and pseudoscalar channels there are large instanton corrections due to direct
interaction of quarks with instantons \cite{Geshkenbein:vb,Novikov:xj}.

In this paper we consider NLO QCD sum rules for pion form factor, where pseudoscalar currents 
are used as pion interpolating currents. As we already mentioned, in this case in contrast to the case of axial
interpolating currents, there is large instanton contribution given by zero-modes of quark
propagators in external (anti-)instanton field. Recently, a lot of quantitative self-consistent
information about instanton properties in QCD was gained from phenomenological estimates and
numerical simulations. For a review of recent advances see \cite{Schafer:1996wv}. As a consequences,
today we are able to take relevant instanton contributions well under control. Another difference
from the conventional approach employing axial currents is the asymptotic of pion electromagnetic
form factor at large momentum transfers. In the pseudoscalar case in limit $Q^2\to\infty$ pion 
electromagnetic form factor scales as $1/Q^4$ \cite{Geshkenbein:jh,Geshkenbein:zs}. There is
simple explanation of this fact\footnote{We are grateful to A.Khodjamirian 
clarifying this point to us}: the well known $1/Q^2$ asymptotic corresponds to the scattering
of virtual photon on the pion in twist 2 state, while taking pseudoscalar currents we artificially 
put pion in twist 3 state. From this discussion, we expect that perturbative contributions
to pion electromagnetic form factor in pseudoscalar case should be of the same order as higher 
twist corrections in the axial case. Our analysis shows, that it is really the case. 
The obtained results of pion form factor calculation are in good agreement with existing experimental data as 
well as with predictions of sum rules based both on pseudoscalar and axial currents. 
There is however discrepancy in the values of different contributions compared to 
previous sum rule analysis based on pseudoscalar currents \cite{Forkel:1994pf}. The source of discrepancy
is discussed in the main body of the paper. While being consistent with the different sum rule 
predictions, our results differ from the predictions made in \cite{Faccioli:2002jd}.      

Besides rich physical content of pseudoscalar channel, which draw our attention to this problem,
we had in mind the development of effective technic for calculation of QCD corrections to double spectral
densities. So, it is natural, that we chose as a playground the simplest situation. 
In our forthcoming publications we are planning to present the details of our method as well
as perform a detail analysis of NLO QCD sum rules with axial interpolating currents. 
Here, we would like to stress that an inclusion of radiative corrections is conceptually very important, 
as only in such a way we can simultaneously account for both hard and soft contributions.
However in present case they turn out to be numerically small.

The paper is organized as follows. In section 2 we describe our framework and give explicit
expressions for next-to-leading order corrections to double spectral density. Next, in section 3
we discuss instanton contributions, supplementing OPE of QCD sum rules in this case. 
Section 4 contains our numerical analysis. Finally, in section 5 we draw our conclusions.

\section{Derivation of QCD sum rules}   

To determine pion electromagnetic form factor we employ the approach of three-point QCD
sum rules. This procedure is similar to that of two-point sum rules and the information
from the latter on the pion coupling to its current is required in order to make
predictions for the pion electromagnetic form factor. Here we are using pseudoscalar interpolating 
currents to describe pion states. The vacuum to pion transition matrix element of pseudoscalar current is 
defined by
\begin{eqnarray}
\langle 0|i\bar u\gamma_5 d|\pi^{-}\rangle = 
f_{\pi}\frac{m_{\pi}^2}{m_u + m_d} = 
-\frac{2}{f_{\pi}}\langle 0|\bar\psi\psi |0\rangle ,
\end{eqnarray}
where $f_{\pi} = 131$ MeV and 
$\left.\langle 0|\bar\psi\psi |0\rangle\right|_{\mu = 1~\mbox{GeV}} = -(0.23 \mbox{GeV})^2$. 
Next, the pion electromagnetic form factor is given by hadronic matrix element
of electromagnetic current:
\begin{eqnarray}
\langle\pi (p')|j_{\mu}^{\mathbf{el}}|\pi (p)\rangle = 
f_{+}(Q^2)(p_{\mu} + p'_{\mu}) + f_{-}(Q^2)(p_{\mu} - p'_{\mu}),
\end{eqnarray}
where $j_{\mu}^{\mathbf{el}} = e_u\bar u\gamma_{\mu}u + e_d\bar d\gamma_{\mu}d$,
the momenta of initial and final state pions were denoted by $p$, $p'$ and $Q^2 = -q^2$ ($q = p -p'$)
is square of momentum transfer. Conservation of electromagnetic current leads to relation
between two form factors introduced above. It is easy to obtain, that
\begin{eqnarray}
f_{-}(Q^2) = f_{+}(Q^2)\frac{s_1 - s_2}{Q^2},
\end{eqnarray} 
where the following notation was introduced: $s_1 = p^2$, $s_2 = p'^2$. 

Following the standard procedure for the evaluation of form factors in the framework of
QCD sum rules, we consider the three-point correlation function:
\begin{eqnarray}
\Pi_{\mu}(p,p',q) = i^2\int dx dy e^{i (p'\cdot x - p\cdot y)}
\langle 0|T\{\bar u(x)\gamma_5 d(x), j_{\mu}^{\mathbf{el}}(0), \bar u(y)\gamma_5 d(y)\rangle
\label{correlator}
\end{eqnarray} 
Similar, to pion form factor decomposition $\Pi_{\mu}(p,p',q)$ is presented as a sum of two
independent Lorentz structures:
\begin{eqnarray}
\Pi_{\mu} = \Pi_{+}(p_{\mu} + p'_{\mu}) + \Pi_{-}(p_{\mu} - p'_{\mu}).
\end{eqnarray}
The scalar amplitudes $\Pi_i$ are the functions of kinematical invariants, i.e.
$\Pi_i = \Pi_i(p^2, p'^2, q^2)$. For the calculation of QCD expression for three-point 
correlator one employs the operator product expansion (OPE) for the $T$-ordered product of currents. As a 
result of OPE one obtains besides leading perturbative contribution also power corrections, given by vacuum
QCD condensates. It turns out, that condensate contributions are small in our case, so 
in what follows we will neglect them. As, we already mentioned, in the case of pseudoscalar
interpolating currents there are essential non-perturbative corrections due to instantons. 
In this section we will consider only perturbative contribution to sum rules, while the
instanton contribution will be considered in next section. 

\begin{figure}[ht]
\begin{center}
\includegraphics[scale=0.5]{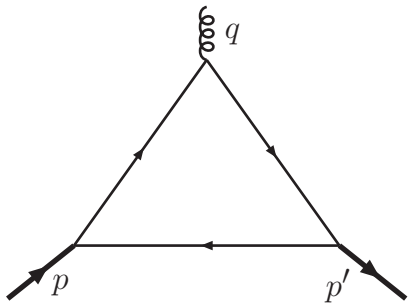} 
\caption{LO diagram}
\label{LOdiagram}
\end{center}
\end{figure}

Next, for the perturbative contribution for each of two scalar amplitudes we write a double dispersion 
relation in variables $s_1 = p^2$ and $s_2 = p'^2$ at $q^2 < 0$:
\begin{eqnarray}
\Pi_i^{\mathbf{pert}}(p^2, p'^2, q^2) = \frac{1}{(2\pi)^2}
\int\frac{\rho_i^{\mathbf{pert}} (s_1,s_2,Q^2)}{(s_1-p^2)(s_2-p'^2)} ds_1ds_2 + \mbox{subtractions}
\label{doubledisp} 
\end{eqnarray} 
The integration region in (\ref{doubledisp}) is determined by condition
\footnote{In the present case this inequality is satisfied identically.}
\begin{eqnarray}
-1 \leq \frac{s_2 - s_1 -q^2}{\lambda^{1/2} (s_1, s_2, q^2)} \leq 1
\end{eqnarray}
and
\begin{eqnarray}
\lambda (x_1, x_2, x_3) = (x_1 + x_2 -x_3)^2 - 4 x_1 x_2.
\end{eqnarray}

\begin{figure}[ht]
\begin{center}
\includegraphics[scale=0.3]{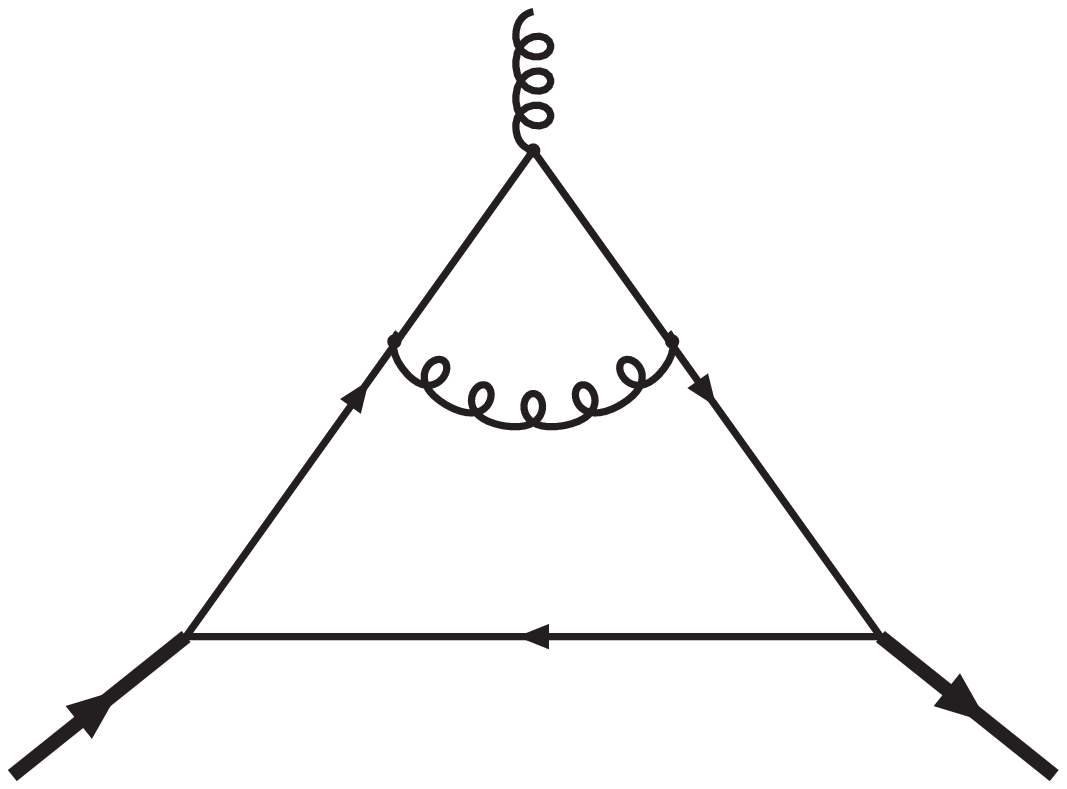} \makebox[2.cm]{}
\includegraphics[scale=0.3]{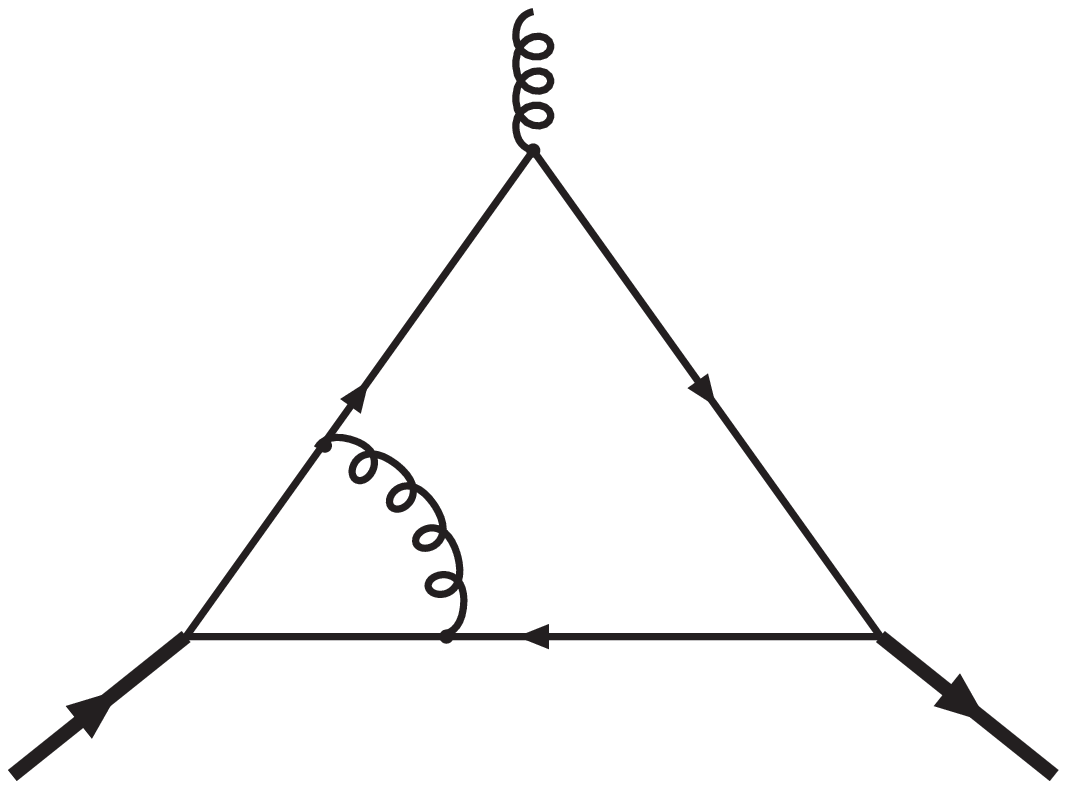} \makebox[2.cm]{} 
\includegraphics[scale=0.3]{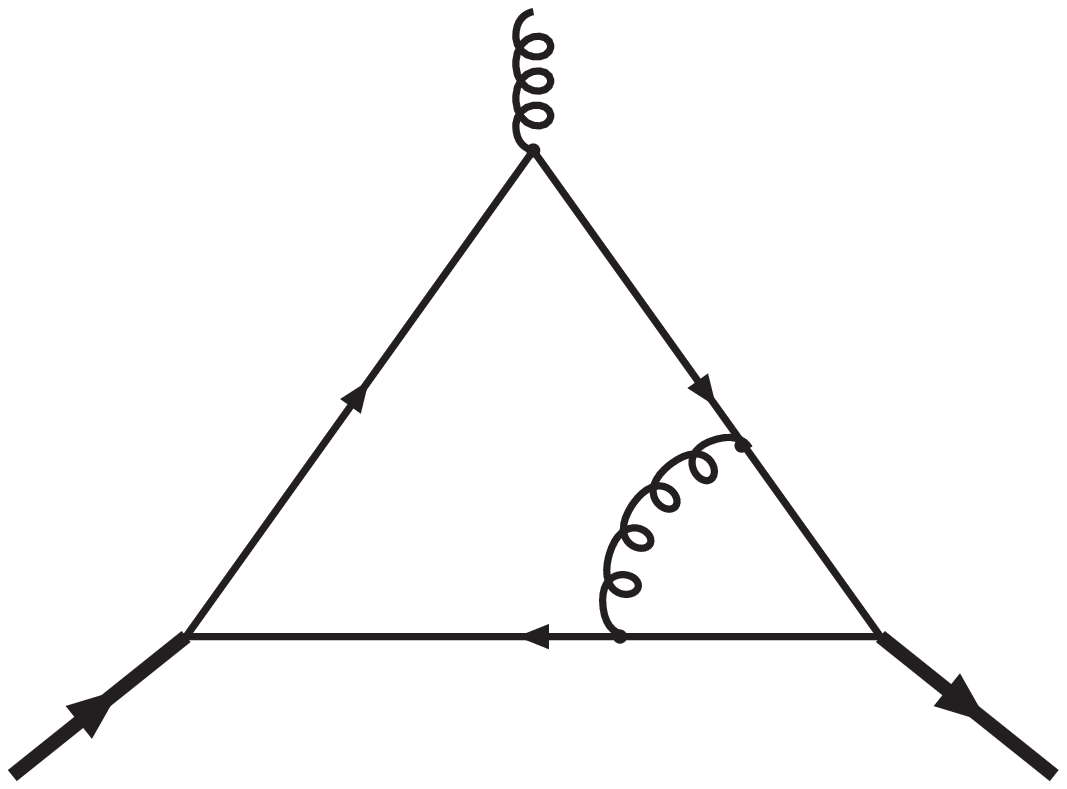} \\ \vspace*{1cm}
\includegraphics[scale=0.3]{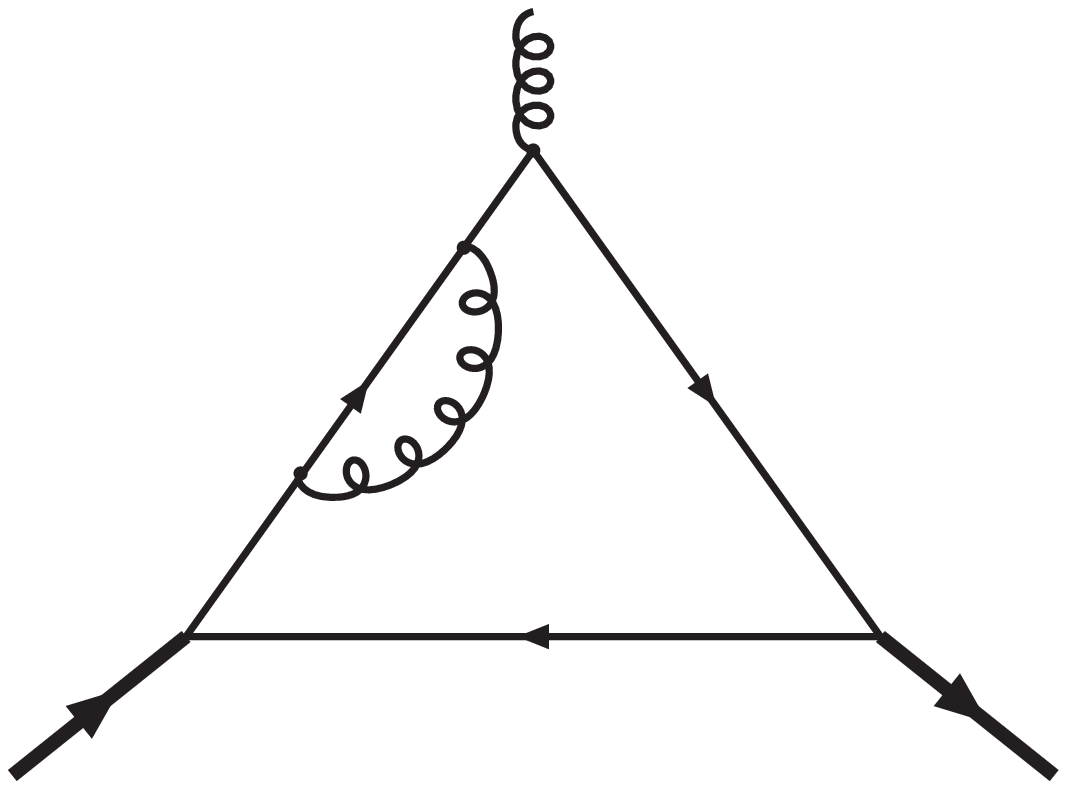} \makebox[2.cm]{}
\includegraphics[scale=0.3]{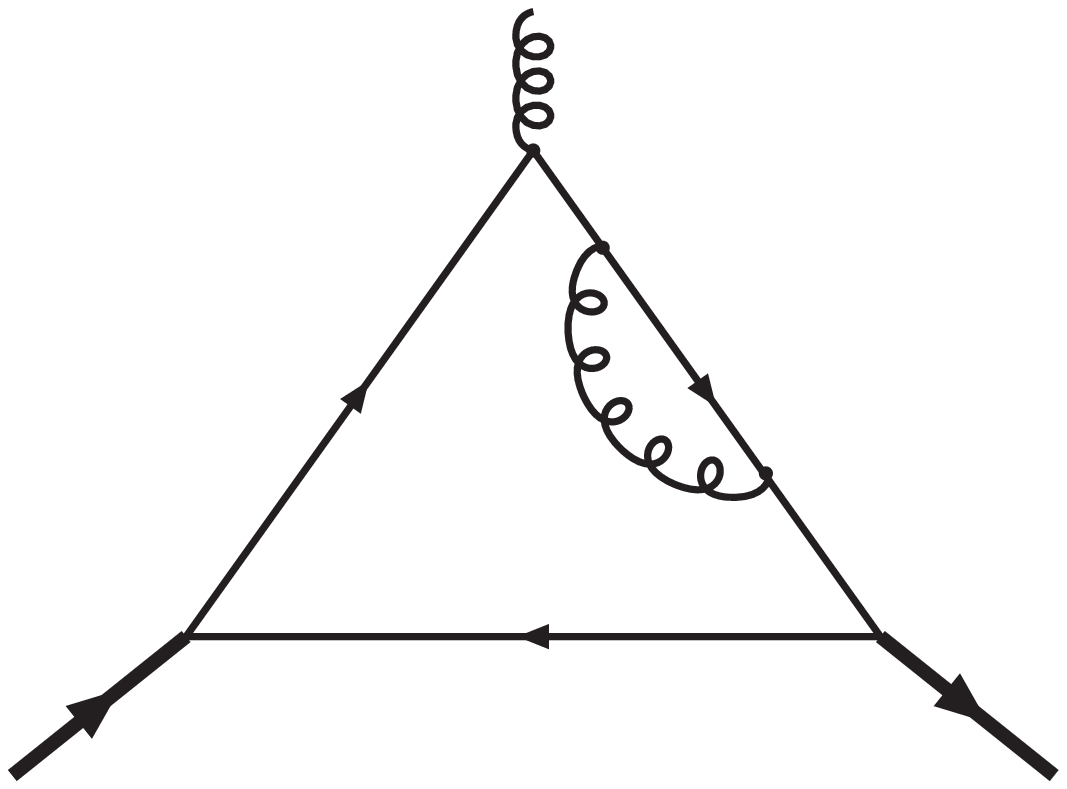} \makebox[2.cm]{} 
\includegraphics[scale=0.3]{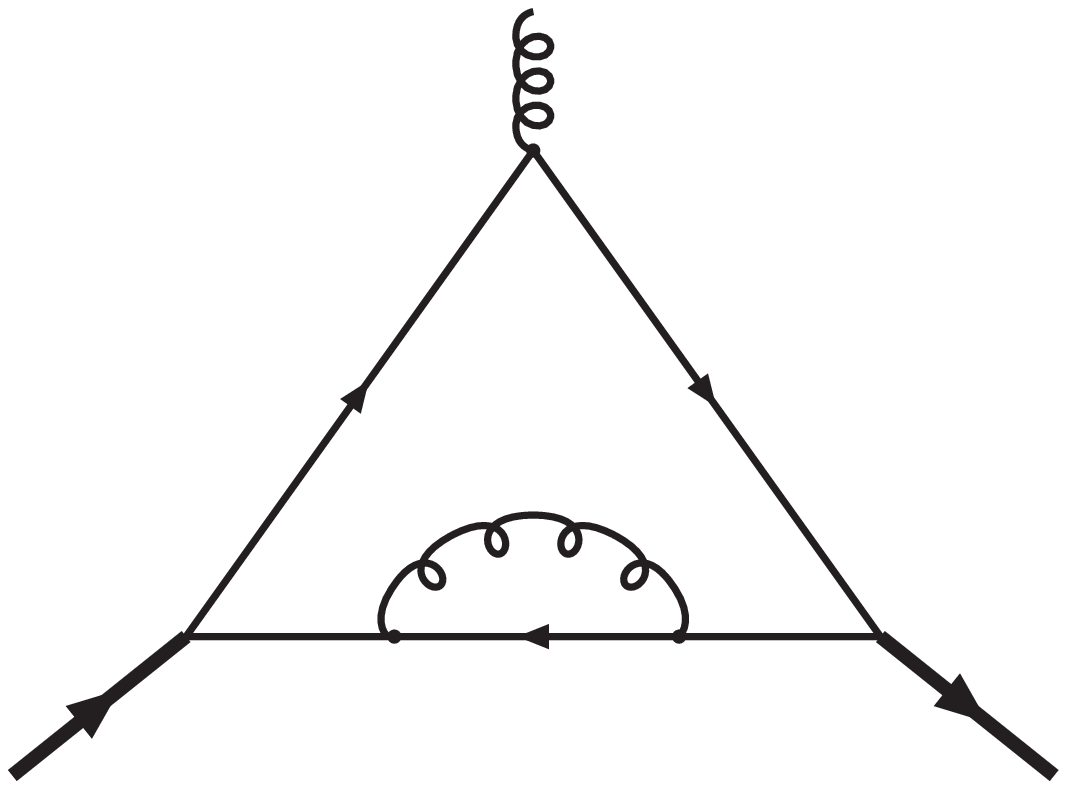}
\vspace*{0.5cm}
\caption{NLO diagrams}
\label{NLO diagrams}
\end{center}
\end{figure}

At leading order in coupling we have only one diagram depicted in Fig. 1, contributing to
three-point correlation function. At next to leading order we have 6 diagrams shown in Fig. 2.
The calculation of corresponding double spectral density was performed with the standard use
of Cutkosky rules. In the kinematic region $q^2 < 0$, we are interested in, there is no 
problem in applying Cutkosky rules for determination of $\rho_i (s_1, s_2, Q^2)$ and integration
limits in $s_1$ and $s_2$. The non-Landau type singularities, not accounted for by Cutkosky
prescription, do not show up here. It is easy to find, that at Born level $\rho_{+}(s_1, s_2, Q^2)$ is
given by:

\begin{eqnarray}
\rho_{+}^{(0)}(s_1, s_2, Q^2) = \frac{3 s_1 s_2 Q^2}{k^{3/2}},
\end{eqnarray}

where $k = \lambda (s_1, s_2, -Q^2)$. Calculation of radiative corrections to double
spectral density is in principle straightforward. One just needs to consider all possible
double cuts of diagrams, shown in Fig. 2. However, the presence of collinear and soft infrared
divergences together with ultraviolet divergences related to renormalization of
pseudoscalar currents calls for appropriate regularization of arising divergences 
at intermediate steps of calculation and makes the whole analytical calculation quite
involved. In this paper we present only the final expression for radiative corrections
to double spectral density, leaving all details of the calculation to separate publication.
We checked, that all infrared divergences cancel, while the remained ultraviolet 
divergences are subtracted with renormalization constants of pseudoscalar currents.

\begin{eqnarray}                
\rho_{+}(s_1, s_2, Q^2) = \rho_{+}^{(0)}(s_1, s_2, Q^2) + 
\left( 
\frac{\alpha_s}{4\pi}
\right)\rho_{+}^{(1)}(s_1, s_2, Q^2),
\end{eqnarray}         

LO spectral density was given before, and for NLO correction we have the following
expression:

\begin{eqnarray}
\rho_{+}^{(1)}(s_1, s_2, Q^2) &=& \frac{4}{3}\frac{Q^2}{k^{3/2}}\cdot \left\{
\right. \nonumber \\
&& - 6s_1s_2\log^2\left(-\frac{x_1}{x_2}\right) - 12s_1s_2\log\left(-\frac{x_1}{x_2}\right)
+ 12s_1s_2\log\left(-\frac{y_1}{s_2}\right)\log\left(-\frac{x_1}{x_2}\right) \nonumber \\
&& + 12s_1s_2\log\left(-\frac{y_1}{y_2}\right)\log\left(-\frac{x_1}{x_2}\right) 
-12s_1s_2\log\left(1-\frac{z_1}{z_2}\right)\log\left(-\frac{x_1}{x_2}\right) \nonumber \\
&& -12s_1s_2\log^2\left(\frac{x_2}{s_1}\right) -12s_1s_2\log^2\left(-\frac{y_1}{s_2}\right)
- 6s_1s_2\log^2\left(-\frac{y_1}{y_2}\right) \nonumber \\
&& + 2\pi^2s_1s_2 + 33s_1s_2 + 9s_1z_1 + 9s_2z_1 + 9Q^2z_1 
-18s_1s_2\log\left(\frac{x_2}{s_1}\right) \nonumber \\
&& + 18s_1s_2\log\left(-\frac{y_1}{s_2}\right) + 
12s_1s_2\log\left(\frac{x_2}{s_1}\right)\log\left(-\frac{y_1}{s_2}\right) 
-6s_1s_2\log\left(-\frac{y_1}{y_2}\right) \nonumber \\
&& -12s_1s_2\log\left(\frac{x_2}{s_1}\right)\log\left(-\frac{y_1}{y_2}\right)
+12s_1s_2\log\left(-\frac{y_1}{y_2}\right)\log\left(1-\frac{z_1}{z_2}\right) \nonumber \\
&& -15s_1s_2\log\left(\frac{z_1}{z_2}\right) - 18s_1s_2\log\left(\frac{s_1^2s_2Q^2}{\mu^8}\right)
+ 18s_1s_2\log\left(\frac{k}{\mu^4}\right) - 12s_1s_2\Li_2\left(\frac{x_2}{s_1}\right) \nonumber \\ 
&& - 12s_1s_2\Li_2\left(-\frac{x_2}{Q^2}\right) + 12s_1s_2\Li_2\left(\frac{x_2}{x_1}\right)
-12s_1s_2\Li_2\left(-\frac{y_1}{s_2}\right) - 12s_1s_2\Li_2\left(\frac{y_1}{Q^2}\right) \nonumber \\
&& +12s_1s_2\Li_2\left(\frac{y_1}{y_2}\right) - 12s_1s_2\Li_2\left(\frac{z_1}{s_1}\right)
-12s_1s_2\Li_2\left(\frac{z_1}{s_2}\right) - 12s_1s_2\Li_2\left(1-\frac{z_2}{z_1}\right) \nonumber \\
&& \left. \right\},
\end{eqnarray}
where the following notation was introduced:
\begin{eqnarray}
x_1 &=& \frac{1}{2}(s_1-s_2-Q^2)-\frac{1}{2}\sqrt{k}, \\
x_2 &=& \frac{1}{2}(s_1-s_2-Q^2)+\frac{1}{2}\sqrt{k}, \\
y_1 &=& \frac{1}{2}(s_1+Q^2-s_2)-\frac{1}{2}\sqrt{k}, \\
y_2 &=& \frac{1}{2}(s_1+Q^2-s_2)+\frac{1}{2}\sqrt{k}, \\
z_1 &=& \frac{1}{2}(s_1+s_2+Q^2)-\frac{1}{2}\sqrt{k}, \\
z_2 &=& \frac{1}{2}(s_1+s_2+Q^2)+\frac{1}{2}\sqrt{k}
\end{eqnarray}
In the limit $Q^2\to\infty$ our NLO double spectral density takes the following asymptotic form:
\begin{eqnarray}
\rho_{+}^{(1)}(s_1, s_2, Q^2) &=& \frac{4}{3}\frac{s_1s_2Q^2}{k^{3/2}}
\left\{
-40\frac{s_1}{Q^2}-40\frac{s_2}{Q^2}-28+26\log\left(\frac{s_1s_2}{\mu^4}\right)
-28\log\left(\frac{Q^2}{\mu^2} \right)
\right\}
\end{eqnarray} 
So, in our case of pseudoscalar currents in contrast to the case of sum rules with axial
interpolating currents and light cone sum rules double logarithms cancel. This result
may be considered strange at first. It is known, that Sudakov factor, suppressing
transverse momenta is required to make pQCD predictions self-consistent at large
momentum transfers\footnote{See, for example \cite{Sterman:1997sx}}. However, as
was already noted in Introduction, in pseudoscalar case taking limit $Q^2\to\infty$ 
we are computing not the leading asymptotic, but higher twist or power correction 
to the leading asymptotic from axial case and thus there is no contradiction to
what we would ordinary expect. Moreover, employing the same technic for the three-point correlation 
function with axial currents\footnote{The axial case will be considered in a separate publication.},
we obtain already known expression for double logarithmic terms.

Now, let us proceed with the physical part of three-point sum rules. The connection
to hadrons in the framework of QCD sum rules is obtained by matching the resulting
QCD expressions of current correlators with spectral representation, derived from
a double dispersion relation at $q^2\leq 0$:
\begin{eqnarray}
\Pi_i(p_1^2, p_2^2, q^2) = \frac{1}{(2\pi)^2}
\int\frac{\rho_i^{\mathbf{phys}}(s_1, s_2, Q^2)}{(s_1 - p_1^2)(s_2 - p_2^2)}ds_1 ds_2
+ \mbox{subtractions}. \label{disp_phys}
\end{eqnarray} 
Assuming that the dispersion relation (\ref{disp_phys}) is well convergent, the physical
spectral functions are generally saturated by the lowest lying hadronic states plus
a continuum starting at some thresholds $s_1^{th}$ and $s_2^{th}$:
\begin{eqnarray}
\rho_i^{\mathbf{phys}}(s_1, s_2, Q^2) &=& \rho_i^{\mathbf{res}}(s_1, s_2, Q^2) + \nonumber \\
&& \theta (s_1 - s_1^{th})\cdot\theta (s_2 - s_2^{th})\cdot\rho_i^{\mathbf{cont}}(s_1, s_2, Q^2),
\end{eqnarray}
where
\begin{eqnarray}
\rho_i^{\mathbf{res}}(s_1, s_2, Q^2) &=& 
\langle 0|\bar d\gamma_5 u|\pi^{-}(p')\rangle 
\langle\pi^{-}(p')|f_i(Q^2)|\pi^{-}(p)\rangle 
\langle\pi^{-}(p)|\bar u\gamma_5 d|0\rangle\cdot \nonumber \\ &&
(2\pi)^2\delta (s_1)\delta (s_2) + \mbox{higher state contributions} 
\end{eqnarray}
In our approximation of massless quarks we put $m_{\pi}^2 = 0$. 
The continuum of higher states is modeled by the perturbative 
absorptive part of $\Pi_i$, i.e. by $\rho_i$. Then, the 
expressions for the form factors $f_i$ can be derived by equating 
the representations for three-point functions $\Pi_i$ from (\ref{doubledisp}) and (\ref{disp_phys}).
This last step constitutes a formulation of QCD sum rules for our particular problem.

\section{Instanton contribution}

Now, let us supplement the conventional operator product expansion in QCD sum rules, presented
in the previous section, with direct instanton contributions. This contribution could be 
taken into account quite accurately using semiclassical approximation, in which all gauge
configurations are replaced by an ensemble of topologically nontrivial fields: instantons and
anti-instantons. Since in the sum rules the correlation function is being probed mainly at
distances $x, y\simeq 0.2~\mbox{fm}$, which are smaller compared to instanton spacing 
$R\simeq 1~\mbox{fm}$, it is natural to use for calculations single instanton approximation (SIA).   
In this approximation the correlation function is dominated by a single instanton, the closest 
one. The effects of other instantons are taken into account at the mean-field level. Another
attractive feature of this approximation is the possibility to carry out all calculations 
analytically. It turns out, that the bulk properties of instanton ensemble could be very well
described with simple parametrization \cite{Shuryak:1982qx}:
\begin{eqnarray}
n (\rho) = \bar n~\delta (\rho - \bar\rho),
\end{eqnarray}
with the average (anti-)instanton density and size are given by
\begin{eqnarray}
\bar n = \frac{1}{2}~\mbox{fm}^{-4},\quad 
\bar\rho = \frac{1}{3}~\mbox{fm}.
\end{eqnarray}
The actual calculation of the closest instanton contribution could be performed with the help
of quark propagator in an instanton background field:
\begin{eqnarray}
S_I (x,y;z) = S_I^{zm}(x,y;z) + S_I^{nzm}(x,y;z).
\end{eqnarray}
The zero-mode part of the propagator is given by \cite{'tHooft:up,'tHooft:fv,Diakonov:1985eg}
\begin{eqnarray}
S_I^{zm}(x,y;z) = \frac{(\hat x -\hat z)\gamma_{\mu}\gamma_{\nu}(\hat y -\hat z)}{8 m}
\left[
\tau_{\mu}^{-}\tau_{\nu}^{+}\frac{1-\gamma_5}{2}
\right]\phi (x-z)\phi (y-z),
\end{eqnarray}
where 
\begin{eqnarray}
\phi (t) = \frac{\rho}{\pi}\frac{1}{|t|(t^2+\rho^2)^{3/2}},\quad 
\tau_{\mu}^{\pm} = (\tau, \mp i).
\end{eqnarray}
and $m$ stands for current quark mass. The corresponding expression in the anti-instanton field is 
obtained through substitutions
\begin{eqnarray}
\frac{1-\gamma_5}{2} \leftrightarrow \frac{1+\gamma_5}{2}, \quad
\tau^{-} \leftrightarrow \tau^{+}
\end{eqnarray} 
In the limit of small distances, which is the case in the present analysis, the expression
for nonzero-mode part of propagator is simplified and is given by
\begin{eqnarray}
S_I^{nzm}(x,y;z) \simeq S_0 (x,y),
\end{eqnarray}
where $S_0$ denotes the free quark propagator. Next, in SIA the effects of other instantons different from 
the closest one are taken care of by substituting the current
quark mass $m$ in the expression for zero-mode part of quark propagator with effective
mass $m^*$. In the simplest approximation, it can be extracted from the value of
quark condensate \cite{Shifman:uw}:
\begin{eqnarray}
m^* = m - \frac{2}{3}\pi^2\rho^2\langle \bar u u\rangle. \label{mquark}
\end{eqnarray}
 
Now, it is quite straightforward to calculate an instanton contribution to
the three-point correlation function (\ref{correlator}), we are interested in. 
It is given by
\begin{eqnarray}
\Pi_{\mu}^{\rm inst}(p,p',q) &=& - \frac{4}{\pi^6}\frac{\bar n\bar\rho^6}{m^{*2}(\bar\rho)}
\int d^4x\int d^4 y\int d^4 z~ e^{ip'\cdot x}e^{-iq\cdot y}e^{i(p'-q)\cdot z}\times \nonumber \\
&& 
\left[
\frac{y^2 (y-z)_{\mu} - (y+z)^2y_{\mu}}{(x^2+\bar\rho^2)^3|y|(y^2+\bar\rho^2)^{3/2}
(y+z)^4|z|(z^2+\bar\rho^2)^{3/2}} +
\pmatrix{ z_{\mu} \leftrightarrow x_{\mu} \cr y_{\mu} \leftrightarrow - y_{\mu}}
\right],
\end{eqnarray}
where the sum over instanton and anti-instanton contributions together
with integration over their positions were performed. In what follows we will
perform the numerical analysis for pion electromagnetic form factor using Borel
scheme. A corresponding Borel transformed Borel instanton contribution was
calculated in \cite{Forkel:1994pf}. The resulting expression could be found in the section with
numerical results after the definition of Borel transformation.

\section{Numerical analysis}

For numerical analysis we are using so called Borel scheme of QCD sum rules. That is, 
to get rid of unknown subtraction terms in (\ref{doubledisp}) let us exploit the
Borel transformation procedure in two variables $s_1$ and $s_2$. We define the Borel
transform of three-point function $\Pi_i (s_1, s_2, q^2)$ as
\begin{eqnarray}
\Phi_i (M_1^2, M_2^2, q^2)&\equiv & \hat B_{12}\Pi_i (s_1, s_2, q^2) =  \nonumber \\ &&
\lim_{n,m\to\infty}\left\{\left. \frac{s_2^{n+1}}{n!}
\left(-\frac{d}{d s_2}\right)^n \frac{s_1^{m+1}}{m!}
\left(-\frac{d}{d s_1}\right)\right|_{s_1=m M_1^2, s_2 = n M_2^2} \right\}
\Pi_i (s_1, s_2, q^2) \nonumber \\  \label{boreltransform}
\end{eqnarray}
Borel transformation (\ref{boreltransform}) of (\ref{doubledisp}) and (\ref{disp_phys}) gives 
\begin{eqnarray}
\Phi_i (M_1^2, M_2^2, q^2) = \frac{1}{(2\pi)^2}\int_0^{\infty}ds_1\int_0^{\infty}ds_2
\exp\left[-\frac{s_1}{M_1^2}-\frac{s_2}{M_2^2}\right]\rho_i (s_1, s_2, q^2), \label{borelcor}
\end{eqnarray}
In the following we put $M_1^2 = M_2^2 = M^2$. If $M^2$ is chosen to be of order 1 GeV$^2$, 
then the right hand side of (\ref{borelcor}) in the case of physical spectral density will
be dominated by the lowest hadronic state contribution, while the higher state contribution 
will be suppressed.

Equating Borel transformed theoretical and physical parts of QCD sum rules we get
\begin{eqnarray}
&& \left(\frac{\alpha_s (\mu_{n})}{\alpha_s (M)} \right)^{-8/9}\left\{
\phi^{(0)}(M^2,Q^2) + \frac{\alpha_s (\mu_n)-\alpha_s (M)}{4\pi}\cdot\frac{1142}{81}
\cdot\phi^{(0)}(M^2,Q^2) \right. \nonumber \\ &&
\left. ~~~~~~~~~~~~~~~~~~~~+\frac{\alpha_s (M)}{4\pi}\phi^{(1)} (M^2, Q^2)  
- \frac{\bar n M^2}{m^{*2}(\rho)}\phi^{\rm inst}(\bar\rho^2Q^2,\bar\rho^2M^2)
\right\} =
\frac{4}{f_{\pi}^2}\langle 0|\bar\psi\psi |0\rangle^2\cdot F_{\pi}(Q^2), \nonumber \\ &&
\end{eqnarray}
where $F_{\pi} (Q^2) = f_{+}(Q^2)$, $\mu_n$ is the normalization point ($\mu_n\simeq 0.5$ GeV). 
We also introduced the following notation:
\begin{eqnarray}
\phi^{(0)}(M^2,Q^2) &=& \frac{1}{(2\pi)^2}\int_0^{s_0} dx\exp\left[-\frac{x}{M^2}\right]
\int_0^x dy \rho_{+}^{(0)} (s_1, s_2, Q^2), \nonumber \\
\phi^{(1)}(M^2,Q^2) &=& \frac{1}{(2\pi)^2}\int_0^{s_0} dx\exp\left[-\frac{x}{M^2}\right]
\int_0^y dy \rho_{+}^{(1)}(s_1, s_2, Q^2),
\end{eqnarray}
where $x = s_1 + s_2$ and $y = s_1 - s_2$. Here, for continuum subtraction we used so called
"triangle" model. To verify the stability of our results with respect to choice of continuum
model we checked, that the usual "square" model gives similar predictions for pion electromagnetic
form factor provided $s_0\sim 1.5 s_1$ is chosen. For more information about different continuum subtraction
models see \cite{Ioffe:qb}. 

\begin{figure}[ht]
\vspace*{3cm}
~~~~~~~~~~~~~~~~$Q^2F_{\pi}(Q^2)$
\vspace*{-3cm}
\begin{center}
\includegraphics[scale=1.]{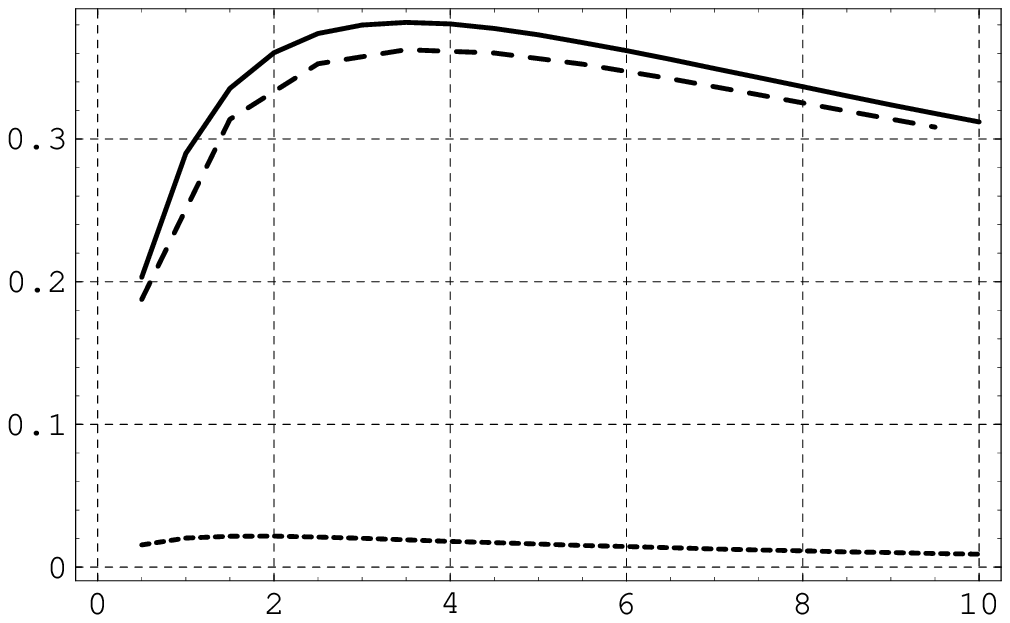} 
\caption{$Q^2$ dependence of pion electromagnetic form factor}
\label{formfactorplotfull}
\end{center}
\vspace*{-1.5cm}
~~~~~~~~~~~~~~~~~~~~~~~~~~~~~~~~~~~~~~~~~~~~~~~~~~~~~~~~~~~~~~~~~~~~~$Q^2$
\vspace*{1.cm}
\end{figure}

\noindent 
The instanton contribution $\phi^{\rm inst}(\bar\rho^2 Q^2,\bar\rho^2 M^2)$ has the following
expression \cite{Forkel:1994pf}:
\begin{eqnarray}
\phi^{\rm inst}(z_1,z_2) &=& \int_0^{\infty}d\alpha \int_0^{\frac{1}{z_2}} d\beta 
e^{-\alpha z_1}e^{-(\omega + \sigma)}e^{-\frac{z_2}{4(1-\beta z_2)}}\cdot
\frac{\alpha\beta}{A^4 (1-\beta z_2)}\times \nonumber \\ &&
\left\{
H(\omega)H(\sigma)\left[
\frac{\alpha + \beta}{z_2}
\left(
\alpha z_1 + \frac{\beta z_2}{16}\frac{z_2 - 8(1-\beta z_2)}{(1-\beta z_2)^2} - 3
\right) \right.\right.\nonumber \\ &&
\left.\left.
-\frac{2\alpha\beta}{A}(2\alpha\beta - A) - \frac{\alpha^3\beta}{4z_2 A^2}
\right]
- \frac{\alpha\beta (\alpha + \beta)}{z_2A}I_1(\omega)H(\sigma)
+ \frac{\alpha^2\beta^2}{A}H(\omega)I_1 (\sigma)
\right\}, \nonumber \\
\end{eqnarray}
where $A = \frac{\alpha + \beta}{z_2} + \alpha\beta$, $\omega = \frac{\alpha}{8A}$, 
$\sigma = \frac{1}{8z_2A}$ and $H(z) = I_1(z) - I_0(z)$ is defined in terms of the
modified Bessel functions $I_n(z)$.

\begin{figure}[ht]
\vspace*{3cm}
~~~~~~~~~~~~~~~~$Q^2F_{\pi}(Q^2)$
\vspace*{-3cm}
\begin{center}
\includegraphics[scale=1.]{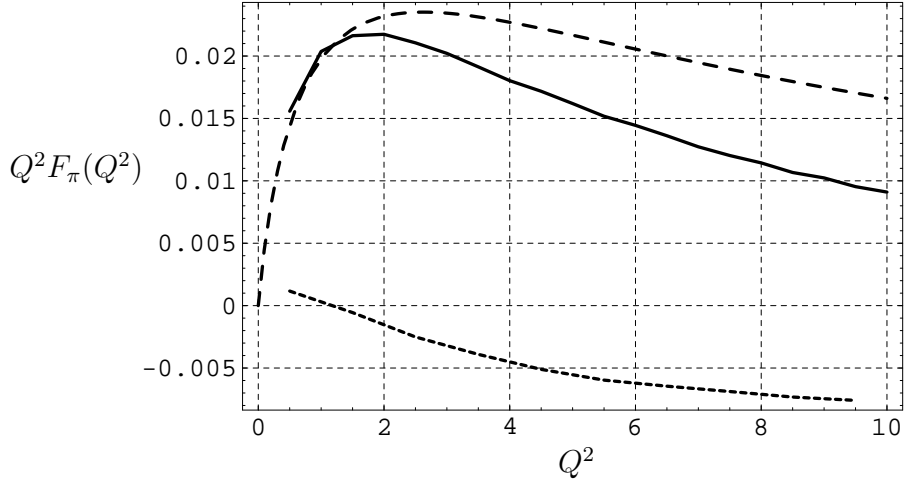} 
\caption{$Q^2$ dependence of perturbative contribution to pion electromagnetic form factor}
\label{formfactorplotpert}
\end{center}
\vspace*{-1.5cm}
~~~~~~~~~~~~~~~~~~~~~~~~~~~~~~~~~~~~~~~~~~~~~~~~~~~~~~~~~~~~~~~~~~~~~$Q^2$
\vspace*{1.cm}
\end{figure}

In the above sum rules for pion electromagnetic form factor we also have taken into account
the evolution of pion pseudoscalar current $J(\mu)$ from scale $M$ down to normalization scale $\mu_n$: 
\begin{eqnarray}
J (\mu_2) = J(\mu_1) U(\mu_1, \mu_2),
\end{eqnarray}
where
\begin{eqnarray}
U(\mu_1, \mu_2) = \left(\frac{\alpha_s (\mu_1)}{\alpha_s (\mu_2)} \right)^{\frac{\gamma_1}{\beta_1}}
\left(1 + \frac{\alpha_s (\mu_1) - \alpha_s (\mu_2)}{4\pi}\frac{\gamma_1}{\beta_1}
\left[\frac{\gamma_2}{\gamma_1} - \frac{\beta_2}{\beta_1} \right]  \right),
\end{eqnarray}
and 
\begin{eqnarray}
\beta_1 &=& -2 \left(11-\frac{2}{3}N_f \right), \\
\beta_2 &=& -4 \left(51 -\frac{19}{3}N_f \right), \\
\gamma_1 &=& -2 (- 3 C_F), \\ 
\gamma_2 &=& -4 \left(
\frac{79}{12}C_FC_A - \frac{11}{6}C_FN_f - \frac{3}{4}C_F^2.
\right)
\end{eqnarray}
Here $C_A = N_c$, $C_F = \frac{N_c^2 -1}{2N_c}$, $N_c$ - number of colors and $N_f$ denotes
number of active flavors. 

\begin{figure}[ht]
\vspace*{3cm}
~~~~~~~~~~~~$F_{\pi}(1~\mbox{GeV}^2)$
\vspace*{-3cm}
\begin{center}
\includegraphics[scale=1.]{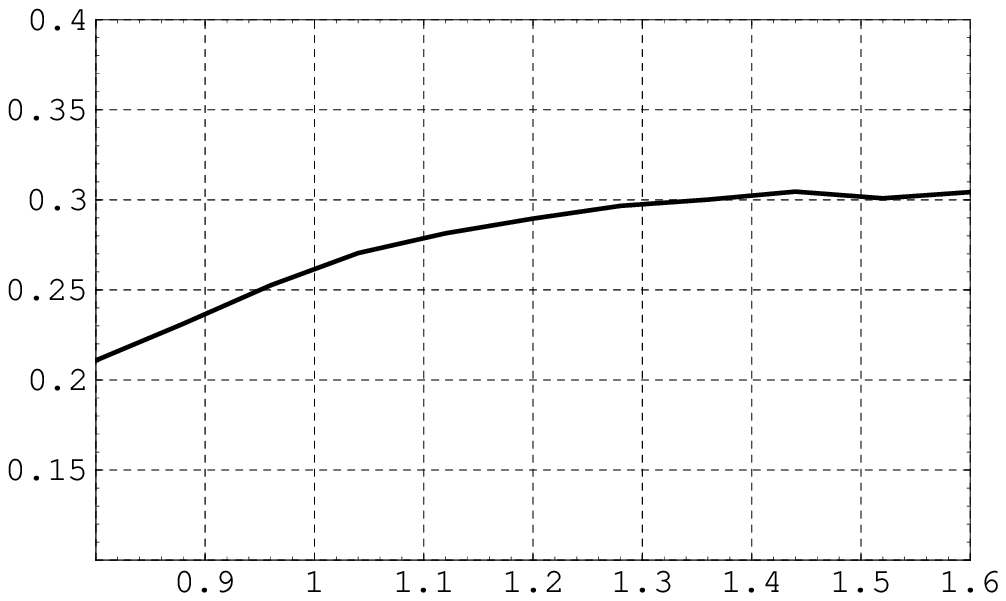} 
\caption{Borel mass $M^2$ dependence of pion electromagnetic form factor at $Q^2 =1~\mbox{GeV}^2$}
\label{formfactorplotM2}
\end{center}
\vspace*{-1.5cm}
~~~~~~~~~~~~~~~~~~~~~~~~~~~~~~~~~~~~~~~~~~~~~~~~~~~~~~~~~~~~~~~~~~~~~$M^2$
\vspace*{1.cm}
\end{figure}

For the continuum threshold we use the value $s_0 = 3.6~\mbox{GeV}^2$ extracted from corresponding two-point
sum rules for pion \cite{Shuryak:1992ke}. The relatively large separation of continuum from lowest
resonance in pseudoscalar channel (compared to vector and axial channels) is explained by almost
complete dominance of the pion at low energies in this particular channel. A few words should be
said about the value of $m^*(\bar\rho)$ used in numerical analysis. The analysis of \cite{Forkel:1994pf}
uses the value extracted from the quark condensate Eq.(\ref{mquark}). Later, a systematic analysis
of SIA \cite{Faccioli:2001ug} showed, that in the instanton vacuum with realistic density the square of effective 
mass parameter $m^{*2}(\bar\rho)$ is not given by the square of effective mass $m^*(\bar\rho)$ extracted
from quark condensate Eq.(\ref{mquark}). The real value of $m^{*2}(\bar\rho)$ is sufficiently smaller.
In our analysis we use $m^{*2}(\bar\rho) = (70~\mbox{MeV})^2$. Next, we use two-loop renormalization 
group running of strong coupling constant with $\Lambda^{[3]}_{\rm QCD} = 325~\mbox{MeV}$. In Fig. \ref{formfactorplotM2}
we plotted the dependence of full electromagnetic pion form factor at fixed $Q^2 = 1~\mbox{GeV}^2$ 
(sum of instanton and perturbative contributions) on the value of Borel parameter $M^2$. We see that
"stability plateau" starts to develop for $M^2 > 1.2~\mbox{GeV}^2$. To determine $Q^2$-dependence of
pion form factor in what follows we fix $M^2 = 1.2~\mbox{GeV}^2$. As a result    
we get the results for pion electromagnetic form factor shown in Fig.\ref{formfactorplotfull}
(solid line is the sum of instanton and perturbative contributions, curve with long dashes denotes instanton contribution
and curve with short dashes stands for perturbative contributions.). For completeness, in Fig.\ref{formfactorplotpert}
we plotted the relative magnitude of LO and NLO perturbative contributions solid line is the sum of LO and 
NLO, curve with long dashes denotes LO contribution and curve with short dashes 
stands for NLO contributions.).

We see, that obtained results for pion electromagnetic form factor are in good agreement with available experimental
data. Our results are also in agreement with the predictions of sum rules employing axial pion currents.
The dominant contributions to pion electromagnetic form factor is given by instantons. The difference
in relative sizes of perturbative and instanton corrections computed here compared to similar quantities
from \cite{Forkel:1994pf} is explained by wrong value of $m^{*2}(\bar\rho)$ used in the latter analysis.
On the other hand, the small value of perturbative contribution is in agreement with our expectations:
it should correspond to higher twist correction to leading twist 2 contribution extracted from sum rules with
axial interpolating currents for pion. Also, our perturbative correction has the same value of magnitude as
similar higher twist contributions considered in the framework of light-cone sum rules in 
\cite{Braun:1999uj,Bijnens:2002mg}. So, we may conclude here that pion electromagnetic form factor
is almost saturated by instanton contribution in pseudoscalar channel. It should be mentioned, that
uncertainties of theoretical predictions in pseudoscalar case are still large (conservatively it is up to 50\%) and are 
mainly due to strong dependence of results on the values of normalization point $\mu_n$, $\Lambda^{[3]}_{\rm QCD}$ 
and square of effective mass $m^{*2}(\bar\rho)$. These uncertainties could be further reduced by considering
simultaneously sum rules for two-point correlation function of pseudoscalar currents.

\section{Conclusion}

We presented the results for pion electromagnetic form factor in the framework of three-point 
NLO QCD sum rules with pseudoscalar interpolating currents for pions. We redone the analysis 
of instanton contribution to three-point correlation function in this case. It was found, 
that while overall prediction for pion electromagnetic form factor made in \cite{Forkel:1994pf}
holds true, the relative magnitude of different contributions is completely different. The
instanton contribution almost saturates the prediction for pion form factor.
The theoretical curve obtained for $Q^2$ dependence of pion form factor is in a good agreement with
existing experimental data. Here, we for the first time computed radiative corrections
to three-point sum rules with pseudoscalar currents. While, these corrections turned out to
be small numerically the technic developed and tested here will certainly have numerous 
applications in determining QCD corrections to the form factors of other light mesons. 
We are planning to present in our forthcoming publications both the details of NLO calculation together 
with the results for pion form factor in a more theoretically clean case of axial interpolating currents.

We would like to thank A.~Khodjamirian for stimulating discussions and critical comments. The work 
of V.B. was supported in part by Russian Foundation of Basic Research under grant 01-02-16585, 
Russian Education Ministry grant E02-31-96 and CRDF grant MO-011-0.  
The work of A.O. was supported by the National Science Foundation under 
grant PHY-0244853 and by the US Department of Energy under grant DE-FG02-96ER41005.

\end{document}